\documentclass[a4paper]{JHEP3}

\usepackage{epsfig,multicol,bbm}

\newcommand{\be}{\begin{equation}}
\newcommand{\ee}{\end{equation}}
\newcommand{\bea}{\begin{eqnarray}}
\newcommand{\eea}{\end{eqnarray}}
\newcommand{\mbb}{\mathbb}
\newcommand{\ti}{\times}

\newcommand{\mc}{\mathcal}
\newcommand{\gsim}{\gtrsim}
\newcommand{\lsim}{\lesssim}

\newcommand{\ma}{m_{\scriptscriptstyle A}}
\newcommand{\mb}{m_{\scriptscriptstyle B}}

\def\dsl{\hbox{/\kern-.5900em$\partial$}}
\def\pref#1{(\ref{#1})}

\title{Continuous Global Symmetries and Hyperweak Interactions
in String Compactifications}

\author{C.P. Burgess$^{\small{1-3}}$, J.P. Conlon$^{4,5}$,  L.-Y. Hung$^4$,
  C.H. Kom$^4$, A. Maharana$^4$, F. Quevedo$^4$
\\$^1$ Perimeter Institute for Theoretical Physics, Waterloo ON, N2L
2Y5, Canada.
\\$^2$ Physics $\&$ Astronomy, McMaster University, Hamilton ON, L8S 4M1,
Canada.
\\$^3$ Theory Division, CERN, CH-1211 Geneva 23, Switzerland.
  \\$^4$ DAMTP, University of Cambridge, Wilberforce
  Road, Cambridge, CB3 0WA, UK.
  \\$^5$ Cavendish Laboratory, J J Thomson Avenue, Cambridge CB3 0HE, UK \\
}
\preprint{DAMTP-2008-45 \\ CAV-HEP/08-08}

\abstract{ We revisit general arguments for the absence of exact
continuous global symmetries in string compactifications and
extend them to D-brane models. We elucidate the various ways
\emph{approximate} continuous global symmetries arise in the
4-dimensional effective action. In addition to two familiar
methods - axionic Peccei-Quinn symmetries and remnant global
abelian symmetries from Green-Schwarz gauge symmetry breaking - we
identify new ways to generate approximate continuous global
symmetries. Two methods stand out, both of which occur for local
brane constructions within the LARGE volume scenario of moduli
stabilisation. The first is the generic existence of continuous
{\it non-abelian} global symmetries associated with local
Calabi-Yau isometries. These symmetries are exact in the
non-compact limit and are spontaneously broken by the LARGE
volume, with breaking effects having phenomenologically
interesting sizes ($\sim 0.01$) for plausible choices for
underlying parameters. Such approximate flavour symmetries are
phenomenologically attractive and may allow the fermion mass
hierarchies to be connected to the electroweak hierarchy via the
large volume. The second is the possible existence of new
hyper-weak gauge interactions under which Standard Model matter is
charged, with $\alpha_{HW} \sim 10^{-9}$. Such groups arise from
branes wrapping bulk cycles and intersecting the local (resolved)
singularity on which the Standard Model is supported. We discuss
experimental bounds for these new gauge bosons and their
interactions with the Standard Model particles. }

\begin{document}
\section{Introduction}

Symmetry principles are one of the deepest and most powerful
concepts in theoretical physics. They underly both the forces and
matter content of the Standard Model, these being nothing more
than the local symmetry group and its representations. The
presence or absence of symmetries determines the decay rate of
many Standard Model particles, and can also be used to give an
analytic handle on the strongly coupled physics of QCD through
chiral perturbation theory. Symmetries are also integral to many
ideas for physics beyond the Standard Model (for example
supersymmetry) and also for understanding the structure of Yukawa
couplings within the Standard Model.

String theory is a powerful set of ideas and is very promising as
a conceptual structure that goes beyond effective field theory by
providing a self-consistent ultraviolet completion.
However the large number of string vacua make it difficult to find
a decisive low-energy way to confront string theory with
experiment. This diversity of vacua does not prevent some model-independent
statements. In particular, while in field theory local and global
symmetries are both equally natural, in string theory this is not so:
global symmetries are in almost all cases gauge symmetries.

In this paper we revisit the arguments for the absence of exact
global symmetries in string compactifications \cite{BanksDixon}.
Our goal is to explore the reach and limitations of these
arguments as a constraint on the low-energy effective actions that
are relevant to phenomenology and cosmology. Owing to the
importance of symmetries for understanding low-energy physics, we
regard this study as timely given the significant recent progress
made in string phenomenology through the understanding of moduli
stabilisation.

As the original argument of \cite{BanksDixon} was made using the
heterotic string, while much current model building involves
D-branes, we first analyse the effects of the open string sector
keeping localised D-brane models in mind. Our main conclusion here
is that, as generally believed, it is the closed string sector
that forbids exact global symmetries and contains the
corresponding gauge field in its massless spectrum. The open
string sector does not by itself require the gauging of global
symmetries and can admit continuous global symmetries. However,
open string theories necessarily contain closed strings and so the
result still holds.

This however opens the possibility of having {\it approximate}
continuous global symmetries, which become exact in the limit
where closed string modes decouple. We argue this can be achieved
in local models of D-branes with very large extra dimensions. To
do so we first enumerate the three possible ways of obtaining
approximate global symmetries at low-energies when the microscopic
theory only admits local symmetries. We then describe two distinct
and new ways that string models can generate approximate global
symmetries.
\begin{enumerate}
\item{} {\it Approximate Isometries:} Although compact Calabi-Yau
metrics have no isometries, non-compact local Calabi-Yau metrics
do. In compact embeddings such local isometries can remain as
approximate symmetries. If the localised Standard Model sector is embedded
into a large bulk, approximate isometries of the localised region
of the Calabi-Yau where the Standard Model resides can manifest
themselves as approximate global (flavour) symmetries of the low-energy
field theory describing the Standard Model interactions.
\item{}
{\it Hyperweak Interactions:}
 Gauge symmetries acting on the matter sector
become indistinguishable from global symmetries in the limit of
vanishing gauge couplings. In local brane constructions with large
bulk volumes, branes that wrap bulk cycles and intersect with
Standard Model matter can give hyperweakly coupled gauge groups
that in the extreme large volume limit represent approximate
global symmetries.
\end{enumerate}

We explore how both possibilities arise in the phenomenologically
appealing LARGE\footnote{The capital `LARGE' emphasises that the
volume is \emph{enormously} large compared to the string scale.}
volume scenario developed in \cite{hepth0502058, hepth0505076}.
This combines $\alpha'$ and non-perturbative corrections in IIB
flux vacua to stabilise the compactification volume at values that
are exponentially large in the string coupling,\footnote{Unless
otherwise specified, throughout this paper numerical values of
lengths and volumes will all be in units of the fundamental
(string) scale, $l_s = 2 \pi \sqrt{\alpha'}$.}
\be
 \mc{V} = \frac{\textrm{Volume}}{l_s^6} \sim e^{{c}/{g_s}},
 \quad \textrm{for $c$ an $\mc{O}(1)$ constant.}
\ee
The volume is thus essentially arbitrary and runs through an
enormous range of values for small changes in the underlying
parameters. The geometry resembles a Swiss cheese: in addition to
the LARGE bulk volume $(\mc{V} \sim \tau_b^{3/2})$ (`size of the
cheese'), there are also small blow-up cycles of size $\tau_s \sim
\ln \mc{V}$ (`holes in the cheese'). Standard Model matter and
interactions are assumed to be supported on branes wrapping a
small cycle.

The resulting LARGE volume is a powerful tool to generate
hierarchies and predicts an interestingly complicated pattern of
mass scales. Phenomenologically the most attractive value is
$\mc{V} \sim 10^{15}$. This brings many of the attractive features
of intermediate scale string scenarios \cite{hepph9809582,
hepph9810535}, such as a solution to the hierarchy problem through
TeV supersymmetry breaking using $ m_{3/2} \sim
\frac{M_P}{\mc{V}}, $ an axion decay constant $f_a \sim 10^{11}
\hbox{GeV}$ in the allowed window \cite{hepth0602233}, and a
potentially natural scale for neutrino-mass generating operators
$\Lambda \sim 10^{14} \hbox{GeV}$ \cite{hepph0611144}. On the IIB
side aspects of this scenario have been studied in
\cite{hepth0605141, hepth0610129, 07040737, 07043403, 07053460,
07070105, 07081873, 07100873, 07113389, 08051029}. The mirror IIA
description of these models was recently constructed in
\cite{08041248}. While much of our discussion will be general,
throughout this paper we picture the Standard Model as residing on
a localised stack of branes within an exponentially large volume
Calabi-Yau. In the context of large extra dimensions, flavour
symmetries have been explored in \cite{nima} in the context of the
ADD model, by postulating a flavour symmetry broken on a brane far
away from the Standard Model brane.

This paper will focus on the conceptual aspects of realising
continuous global symmetries in string theory. We will not try and
construct specific singularities and brane configurations
realising either the MSSM or some extension of it, nor will we
discuss the bounds that FCNCs may pose for concrete models of
flavour in the MSSM. One reason for this is that it is far from
clear what field theory and matter content we should aim at -
there are no strong theoretical reasons why there should not be
either new gauge groups or new exotic matter at the TeV scale.
Examples of papers focusing on the model-building aspects of local
constructions are \cite{hepth0005067, hepth0508089, hepth0610007,
08023391}. We instead look for generic features which we may hope
to be phenomenologically relevant even if any particular model is
not.

The paper is organised as follows. We start by enumerating in
section 2 the three ways in which a global symmetry can emerge in
the low-energy limit of a microscopic theory which has only local
invariance. We then review and extend the Banks-Dixon argument for
the non-existence of continuous global symmetries, including the
effects of the open string sector. Since this latter discussion
uses the language of worldsheet conformal field theory, some
readers may wish to omit it. Our main results are in sections 3
and 4. In section 3 we describe how local metric isometries can
generate \emph{approximate} continuous (non)-Abelian global
symmetries for local brane constructions in Calabi-Yau
compactifications. We describe how the quality of the symmetry is
set by the size of the volume and relate our analysis to more
phenomenological discussions of flavour symmetries. Section 4 is
dedicated to the existence of hyperweak gauge fields and
their mixing with the Standard Model $U(1)$, for which we also
summarise some of the experimental bounds.

\section{General Arguments}

This section starts by describing the general mechanisms whereby
global symmetries can appear to arise in the low-energy limit of
UV completions having only local symmetries. It then outlines the
arguments for the absence of global symmetries in closed-string
theory, together with their extensions to include the open-string
sector.

\subsection{Obtaining Approximate Low Energy Global Symmetries}

There are three ways through which (approximate) global symmetries
may arise within the low energy limit of a microscopic theory
which has none. On the one hand, it can happen that the low energy
symmetry cannot be extended in any way to be also a symmetry of
its full UV completion. Alternatively, if such an extension does
exist the symmetry must be a local one for which there is a gauge
boson in the spectrum. This case then subdivides into two further
cases, depending on whether or not this gauge boson is lighter or
heavier than the cutoff of the low-energy theory. These
considerations lead to the following three categories.

\medskip\noindent{\sl 1.\ Emergent Symmetries:} Symmetries are
emergent if they arise as approximate symmetries in the low energy
limit, but cannot be extended to a symmetry once the full UV
particle content is included. Such symmetries (also known as
`accidental' symmetries \cite{Weinberg}), can arise when the gauge
invariance and particle content of the low-energy theory restrict
the possible low-energy interactions so severely as to
automatically ensure invariance under additional global
symmetries. For instance, conservation of baryon and lepton number
would automatically emerge in this way for any string vacuum whose
low-energy particle content consists purely of the Standard Model
particle content, even if this vacuum did not contain any
combination of $B$ or $L$ as gauge symmetries.

\medskip\noindent{\sl 2.\ Gauge Symmetries with Heavy Gauge Bosons:}
Under certain circumstances a local symmetry can resemble a global
symmetry at low energies if it is spontaneously broken, with the
mass of the corresponding gauge boson being too heavy to allow it
to be included in the low-energy effective theory. Perhaps the
simplest example consists of a $U(1)$ gauge boson, $A_\mu$, coupled
to matter fields, $\psi$ of charge $q$, as well as a
Stueckelberg field, $\sigma$, describing the would-be Goldstone
mode whose consumption by the gauge field gives it its mass, $M$.
The lagrangian for this system is constructed from the two gauge
invariant combinations, $V_\mu = A_\mu - \partial_\mu \sigma$, and
$\chi = e^{-iq\sigma}\psi$, such as
\be
    {\cal L} = - \frac14 \, F_{\mu\nu} F^{\mu\nu} - \frac{M^2}{2}
    V_\mu V^\mu + {\cal L}_m(\chi) + e j^\mu(\chi) V_\mu + \cdots \,,
\ee
with $F_{\mu\nu} = \partial_\mu V_\nu - \partial_\nu V_\mu =
\partial_\mu A_\nu - \partial_\nu A_\mu$, and the local
symmetry allows the gauge choice $\sigma = 0$. The
$\chi$-dependent quantities, ${\cal L}_m$ and $j^\mu$, appearing
here are unconstrained by the local symmetry, and so in general
can be arbitrary and in particular need not be $U(1)$ symmetric.
However it can happen that their lowest-dimension contribution
(which dominates at low energy) can nonetheless have such a
symmetry, such as if the only matter field is described by a Dirac
spinor, for which the lowest-dimension terms in both are ${\cal
L}_m = - m \, \overline \chi \,\chi - \tilde m \, \overline \chi
\chi^c - \overline\chi \dsl \chi + \cdots$. This enjoys the
accidental symmetry $\delta\chi = i \omega  \chi$ if $\tilde m=0$,
such as might perhaps be ensured by a discrete symmetry.
Superconductors provide examples along these lines, for which
spontaneous breaking of electromagnetic gauge invariance does not
imply electric charge non-conservation.

\medskip\noindent{\sl 3.\ Gauge symmetries with Light Gauge Bosons:}
It can also happen that the gauge boson of the local symmetry is
lighter than the cutoff, but is so weakly coupled that its
presence could go unremarked in the low-energy theory. This
resembles a global symmetry because gauge symmetries acting on the
matter sector become indistinguishable from global symmetries in
the limit of vanishing gauge couplings. So one might expect to
find string vacua having approximate global symmetries acting on
particular sectors of the low energy theory, provided the spectrum
also includes very weakly coupled, light spin-one gauge bosons.

\subsubsection*{Previous Examples from String Theory}

Despite the absence of continuous global symmetries in string
theory, there are some well known ways that exploit the above
options to obtain approximate continuous global symmetries of the
low-energy effective field theory.

\begin{enumerate}

\item{} {\it Peccei-Quinn symmetries:} The massless spectrum of
all string theories include RR and/or NS-NS forms that lead to
axion fields in the 4D effective field theory. These resemble
Stueckelberg fields, with a corresponding shift symmetry. This is
a perturbative abelian symmetry and is broken by non-perturbative
effects.

\item{} {\it Nonlinearly Realised Abelian Gauge Symmetries:}
String theory also provides examples of gauge bosons that are
massive at the string scale and with abelian global symmetries
below the scale of the gauge boson mass. The low-energy
description in this case involves coupling the electromagnetic
gauge boson to a (Stueckelberg) Goldstone boson, $\sigma$, as
above. In string models the origin of these low-energy abelian
global symmetries is usually the $B\wedge F$ Green-Schwarz
coupling induced by anomalous or non-anomalous $U(1)$'s, which
upon dualisation induces the $V_\mu V^\mu$ coupling above. The
corresponding $U(1)$ becomes massive by the Stueckelberg mechanism
without necessarily having a Higgs field to break the symmetry.
The corresponding $U(1)$ therefore survives in the low energy
effective theory as a remnant global symmetry. As with the
Peccei-Quinn symmetry, this symmetry is also broken by
non-perturbative effects. In brane models examples of these
$U(1)$s can be combinations of baryon and lepton numbers.

\end{enumerate}

\subsection{Continuous Global and Gauge Symmetries in String Theory}

It is well-known that except in very special circumstances, in string theory
no continuous global symmetries exist and
all spacetime symmetries are gauged \cite{BanksDixon}. We now
review the closed string arguments presented in
\cite{BanksDixon, Polchinski:1998}\footnote{The proof in \cite{BanksDixon}
assumes that the Noether current does not contain world-sheet
ghosts fields. This assumption is removed in the discussion in
\cite{Polchinski:1998}.} and discuss their extension to open
strings.

\subsubsection{Closed strings}

In string theory spacetime symmetries necessarily originates from
world-sheet symmetries, which then have a Noether
current. Consider the case where the holomorphic and
anti-holomorphic parts of the Noether currents are independently
conserved. The symmetry comes in a pair and the corresponding
generator of the world-sheet symmetry has the form
\begin{equation}\label{charges}
    Q = \frac{1}{2\pi i} \oint \Bigl[
    dz j(z) - d\bar{z} \bar{j}(\bar{z}) \Bigr],
\end{equation}
assuming that we have integrated over the superspace Grassmann
coordinates in a superstring theory. Since $Q$ generates a
physical symmetry, it must commute with the BRST charge and
additionally be conformally invariant, implying $j$ and $\bar{j}$
must transform as conformal fields with conformal dimensions
$(1,0)$ and $(0,1)$. Consider for concreteness bosonic string
theory\footnote{In a superstring theory we could replace $\partial
X^{\mu}$ by $\psi^{\mu}e^{-\phi}$, where $\psi^{\mu}$ are the
world-sheet fermions and $\phi$ the super-ghost and the discussion
would follow.} and the pair
\begin{equation}
\label{voperators}
    V= \int d^2z \, j \, \bar{\partial}X^\mu  \exp{(ik.X)},
    \qquad  \bar{V}= \int
    d^2z \, \bar{j} \, \partial X^{\mu}  \exp{(ik.X)}.
\end{equation}
The integrands have conformal dimension $(1,1)$ for $k^2 = 0$.
Both $V$ and $\bar{V}$ are conformally invariant, and inherit BRST
invariance from $j$ and $\bar{j}$.  They are thus a pair of valid
vertex operators corresponding to massless vectors which gauge the
pair of symmetries.

There are two exceptions to the above discussion. The first
concerns axionic symmetries. All worldsheet fields are uncharged
under these and so the charge $Q$ vanishes. In this case the
axionic symmetry exists as a perturbative symmetry of the low
energy action, broken non-perturbatively by instantons, under
which matter fields do not transform.

A second exception is the Lorentz symmetry of uncompactified
space-time. The holomorphic and anti-holomorphic parts of the
corresponding world-sheet Noether currents are not conserved
independently. This stems from the non-compact nature of
spacetime, which gives rise to an infrared divergence in the
Green's function. The Ward identity
\begin{equation}
    \bar{\partial}\langle T(z)j(w)j(0)\rangle = \delta^2
    (z-w)(z\partial_z +1) \langle j(z)j(0) \rangle + \delta^2(z)
    \langle j(z)j(0) \rangle
\end{equation}
cannot be integrated \cite{BanksDixon}. In other words, the
corresponding $j$ and $\bar{j}$ do not transform as conformal
fields and the resultant $V$ and $\bar{V}$ are not valid vertex
operators. The existence of the global Lorentz symmetry is
therefore tied to the presence of a non-compact target space in
the sigma model.

\subsubsection{Open strings in a general system of D-branes}
Open strings are described by boundary conformal field theories
(for a review see for example \cite{Cardy:2004hm}). The boundary
is a collection of points where the open string ends and the
conformal boundary condition is given by
\begin{equation}\label{bc}
    T_{ab} n^a t^b = 0,
\end{equation}
where $T_{ab}$ is the world-sheet energy-momentum tensor, and
$n^a, t^b$ are world-sheet vectors that are respectively parallel
and transverse to the boundary. The boundary condition states that
the off-diagonal component of $T_{ab}$, with mixed parallel and
perpendicular indices, should vanish. This requirement does not
distinguish Neumann or Dirichlet boundary conditions. For a
Neumann boundary, for example, the condition (\ref{bc}) implies
that there is no momentum flowing across the boundary, as is
expected of a free end of an open string. It is convenient to work
with Euclidean world-sheets and define complex coordinates $z,
\bar{z}$. Using world-sheet re-parametrization and conformal
invariance, the boundary can be generally taken to be the real
axis, and the CFT defined only in the upper half-plane. The
doubling trick can be applied\cite{Polchinski:1998,
  Cardy:2004hm}, which essentially means that the (anti)-holomorphic
modes are reflected off the boundary.  The important difference with
closed strings is that open string vertex operators are inserted at
the boundary. Therefore the integral is one dimensional.  In our
coordinates it takes the form
%
\begin{equation}
    V_{{open}} = \int_{{Im}(z) = 0} dz \; V(z).
\end{equation}

Conformal invariance of $V_{\textrm{open}}$ requires that $V(z)$
has conformal dimension $(1,0)$. An open string vertex operator is
thus half of a closed string vertex operator. Any operator that
takes the form
\begin{equation}
    A(z) = \partial X^{\mu} : \exp{(ik.X)}:
\end{equation}
therefore already has conformal dimension $(1,0)$ for $k^2 =0$,
and there is no room for the insertion of extra world-sheet
currents as in (\ref{voperators}). The rotational symmetry of the
Chan-Paton factors does not have world-sheet currents - in fact,
this is not a symmetry of the world-sheet as the open string
  end-points are charged under the Chan-Paton rotation and are shifted around.
We can construct a gauge field vertex operator by attaching the
Chan-Paton labels, which do not carry conformal dimension, to $A$,
\begin{equation}\label{gauge_vertex}
    V_{{gauge}}(z) = \lambda_{\alpha \beta} A(z).
\end{equation}
For other world-sheet symmetries with non-trivial local
world-sheet currents, the charges and currents have the same
properties as in (\ref{charges}) and the currents $j$ must have
conformal dimension $(1,0)$ and be BRST invariant.  These
symmetries are not gauged in the open string sector. An example is
the rotational symmetries of the transverse directions of a
D-brane, which are ungauged independent of whether or not these
directions are compactified.  This rotational symmetry is however
shared with the closed strings which can supply the gauge fields
if the directions are compact.

One slightly different case also deserves a comment. As each open
string has two end-points, there are up to two boundaries that
satisfy different boundary conditions. These can be taken as the
positive and negative real axis, with the boundary changing vertex
operator $V_{BC}$  inserted at the origin. This scenario
corresponds to open strings connecting different D-branes. The
previous discussion concerning open strings ending on the same
brane will continue to apply, except that now the insertion of
$V_{BC}$ in every vertex operator takes up further conformal
dimension (see \cite{Hashimoto:1996he} for explicit examples with
intersecting branes). It is now not even possible to form a
Lorentz vector by inserting $A$ as in (\ref{gauge_vertex}), and
these open strings describe matter fields transforming in
bi-fundamental representations $(\bf{P}, \bf{\bar{Q}})$.

To summarise, world-sheet symmetries that give rise to
non-trivial local world-sheet currents are not gauged in the open
string sector. In fact the presence of the boundary reduces the
symmetry in the open string sector compared to the closed string
sector, as e.g. D-branes break translational symmetries. Since
boundary conditions do not affect local properties of the
world-sheet, these symmetries are automatically shared with closed
strings, which can supply the gauge fields. In the low energy
limit where closed strings decouple, these symmetries become
approximately global - for example the $SU(4)$ R-symmetry on the
world-volume theory of D3 branes.

We conclude that in string theory the closed string sector is the
source of massless vectors that gauge continuous global
symmetries. This fits well with the standard claim that
gravitational theories naturally break continuous global
symmetries as the corresponding charges are not conserved by the
process of black hole evaporation. As it is the closed string
sector that gives gravity it is only this sector that is expected
to forbid global symmetries.

Local D-brane models can have approximate global symmetries, and
these are broken only by the interaction of the branes with the
geometry of the extra dimensions. As the extra dimensions become
very large and the bulk decouples, the quality of the symmetry
increases. In the limit that the bulk decompactifies, these
approximate symmetries take the same status as the Lorentz
symmetries of 4-dimensional spacetime and become exact.

\subsection*{Remarks on non-CFT constructions}

The above argument relies on world-sheet CFT techniques
appropriate to weakly coupled string theory. To discuss the
strongly coupled regimes, one would have to use string theory
duality relations, and non-CFT models such as M and F-theory.
Internal structures and symmetries of the lower dimensional
theories are encoded as geometrical data in M/F theory, by
specifying the compactification manifolds. Symmetries are realised
as symmetries of the manifolds.  While we have far less
understanding and control of M/F theories, it is generally
believed that they are gravity theories in the low energy
limit. Therefore, at least in the low energy limit
where concrete statements can be made about these theories, the
geometrical symmetries are gauged by gravity. In the limit where
gravity is essentially decoupled,\footnote{This is the regime
discussed in recent F-theory constructions of quasi-realistic
models - see for example recent discussions in
\cite{08023391,Donagi:2008ca}.} these symmetries remain as
approximate global symmetries as in perturbative string theory.

\section{Continuous Global Flavour Symmetries}

One of the main phenomenological applications of flavour
symmetries is to the problem of understanding the masses and
mixings of the Standard Model fermions. The fermion masses,
plotted in figure \ref{massplot}, are strongly hierarchical with
inter-generational mass ratios approximately $\mc{O}(100):1$. This
structure strongly suggests a deeper organising principle, and
arguably the most appealing idea is the existence of approximate
flavour symmetries under which Standard Model is charged.
\FIGURE[r]{
\includegraphics[width=6cm]{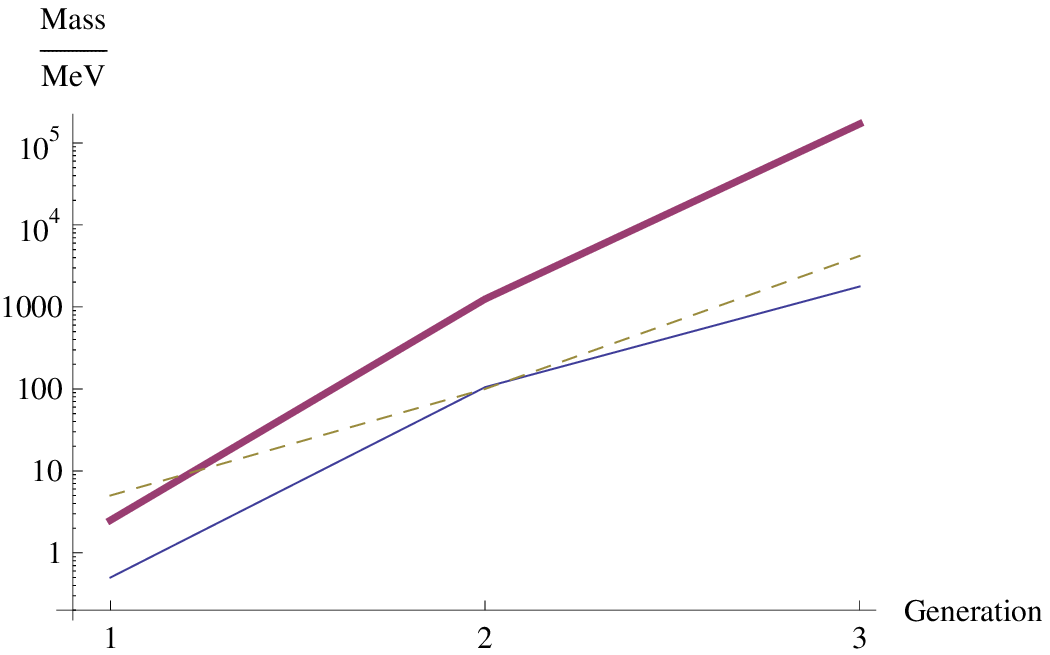}
\caption{The Standard Model fermion masses (excepting neutrinos)
by generation. The thick line represents up type quarks, the
dashed line down type quarks and the ordinary line leptons.
\label{massplot}}}

The previous section has reviewed some ways to obtain approximate
global symmetries. However these methods cannot generate realistic
flavour models: Standard Model matter is uncharged under the
axionic symmetries and in brane models bifundamental quark
doublets must have identical charges under anomalous $U(1)$s. This
is because the $U(1)$ charges of the quark doublets are fixed by
the branes they are connect, which are identical, and so the three
generations necessarily have the same $U(1)$ charge assignments.
This is too restrictive for obtaining realistic Yukawa textures,
using (for example) the Froggatt-Nielsen mechanism
\cite{FroggattNielsen}.

The arguments excluding exact global symmetries do not preclude
the existence of approximate global symmetries. For local brane
constructions of the Standard Model - this is necessarily the case
within the LARGE volume models - physics is determined by the
local geometry and metric. From the viewpoint of the local model,
the metric is that of a non-compact Calabi-Yau. There are many
known examples of such metrics and they often have isometries.
This is familiar from the AdS/CFT correspondence, where isometries
of the local metric corresponds to global flavour or R-symmetries
in the field theory. Examples of local Calabi-Yau metrics and
their isometries are:
\begin{enumerate}
\item
$
\textrm{The local geometry } \mbb{C}^3 \textrm{ with isometry } SO(6),
$
\item
$
\textrm{The conifold, } \sum_i z_i^2 = 0 \textrm{ with local isometry } SU(2) \ti SU(2) \ti U(1) \cite{PhilipXenia},
$
\item
$
\textrm{The resolved } \mbb{C}^3 / \mbb{Z}_3 \textrm{ singularity, }
\mc{O}_{\mbb{P}^2}(-3) \textrm{ with local isometry } SU(3)/\mbb{Z}_3 \cite{FreedmanGibbons, Lutken}.
$
\end{enumerate}

For local brane models embedded in these geometries, isometries
correspond to flavour symmetries. For D3/D7 magnetised brane
models we can view flavour as arising from different solutions of
the Dirac equation. Isometries rotate solutions of the Dirac
equation into each other thereby acting as a flavour symmetry. We
note that such symmetries can only hold for local models, as
global models rely on the whole Calabi-Yau and compact Calabi-Yau
metrics have no continuous isometries. Even for local models, the local
isometry is not expected to survive if the size of the compact
space is $\mc{O}(1)$ in string units - this includes conventional
GUT unification models - as the local metric will then receive
large corrections from the bulk. However in a non-compact
limit,\footnote{Here we refer to a limit in which the scales of
the local geometry are left unaltered while the bulk volume is
taken to infinity.} the local metric is exact and the field theory
has an exact global flavour symmetry.

This is attractive for LARGE volume, as in this case the full
local metric can be sensibly viewed as a perturbation on the
non-compact result and the exact isometry of the non-compact case
will survive as an approximate isometry of the LARGE volume limit.
As it is generic for local Calabi-Yau metrics to have continuous
isometries, this implies that in the context of the LARGE volume
scenario, one should generically expect the low-energy matter
couplings to be governed by approximate continuous flavour
symmetries. We stress that this argument says nothing about the
exact nature of the local gauge group or matter content, or
precisely how the flavour symmetries act. These depend on the
details of the singularity - different singularities have
different isometries - and the local brane construction.

The flavour symmetries act as selection rules on the
superpotential. This is familiar from the AdS/CFT correspondence
and the study of branes at singularities. As a well known example,
the theory of D3 branes on the conifold has a global $SU(2) \ti
SU(2) \ti U(1)$ symmetry \cite{hepth9807080}. In cases where the
bulk is compact, the flavour symmetry ceases to be exact. This
will manifest itself as corrections to the K\"ahler potential (the
flavour symmetry breaking is carried by the K\"ahler moduli which
cannot appear in the perturbative superpotential). We can estimate
the size of the breaking parameter. In local brane models two
scales parametrise the geometry, the length scale of the local
metric ($R_s$) and the length scale of the bulk ($R_b$). These are
related to the sizes of 4-cycles by $\tau_s = R_s^4$ and $\tau_b =
R_b^4$.\footnote{As we are interested in flux compactifications
involving D7 branes, it is 4-cycles that are the natural unit in
which to describe the K\"ahler moduli.} As stringy effects will
modify the classical geometry at small volume, we impose a cutoff
$R_s \ge l_s$ on the effective local length scale. A dimensionless
measure of breaking is set by the ratio of local and global length
scales, ${R_s}/{R_b}$. Roughly, $R_b$ is the distance one needs to
go to feel the bulk metric, and so $R_s/R_b$ measures the effect
of the bulk on the local geometry. Global rescalings of the
metric, $g_{i \bar{j}} \to \lambda^2 g_{i \bar{j}}$, leave
solutions to the Dirac equation unaltered and thus do not modify
the flavour structure. The order parameter for flavour symmetry
breaking is then ${R_s}/{R_b}$.\footnote{We note however that it
is unclear what power of the expansion parameter $\left(
{R_s}/{R_b} \right)$ is appropriate for the Yukawa couplings.}

In phenomenological models based on the LARGE volume scenario, the
size of the overall volume generates the weak scale using $m_{3/2}
= M_P W_0/\mc{V}$. Assuming that the flux superpotential $W_0 \sim
1$,
\be
    \mc{V} \sim \frac{R_b^6}{l_s^6} \sim 10^{14}.
\ee
In contrast the small cycle supports the Standard Model and has a
size $\tau_s \sim R_s^4 \sim 20 l_s^4$. As the bulk radius is $R_b
\sim 100 l_s$ and the local radius $R_s \sim l_s$, the expansion
parameter for the breaking of local isometries is ${R_s}/{R_b}
\sim 0.01$. While the absence of any explicit construction of a
local Standard Model configuration precludes more detailed
analysis, such an expansion parameter is not inconsistent with the
observed pattern of fermion masses.

An attractive feature of this scenario is that it may allow the
fermion mass hierarchies of the Standard Model to be generated by
the same physics, namely the exponentially large volume, that is
used to generate the weak hierachy. The approximate flavour
symmetry is geometric in origin and comes from the same geometric
feature - LARGE volume - used to generate the weak hierarchy. Note
that this is somewhat similar to what occurs in phenomenological
models of Randall-Sundrum scenarios with fermions in the bulk
\cite{hepph0003129}, where extra-dimensional fermion profiles are
separated using the same warping that generated the weak scale
hierarchy.

We finally relate our analysis to the discussion of
\cite{BanksDixon} and section 2 of global flavour symmetries in
string theory. As seen above, the proof that string
compactifications have no continuous global flavour symmetries
relies on the compactness of the internal 2d worldsheet conformal
field theory, and specifically on the fact that the spectrum of
operator conformal dimensions is bounded from below. If the
spectrum is non-compact (for example Minkowksi space time) then
global symmetries can and do exist (for example the Lorentz
group). However, and as described in \cite{BanksDixon}, in cases
where the radius of the internal CFT is much larger than the
string scale \emph{approximate} continuous global symmetries can
exist as to the sigma model the space looks locally non-compact.
For the weakly coupled heterotic string then in vogue, this
large-radius limit is incompatible with a viable phenomenology.
However, this limit is perfectly viable with the advent of
D-branes and local brane constructions.

\subsection*{Connection to Phenomenological Discussions}

The above discussion of flavour symmetries has been geometric and
we would like to connect it to the phenomenological literature
\cite{FroggattNielsen, FlavourSymmetry,FlavourProblem}. One approach
\cite{FroggattNielsen} starts by postulating a new gauged flavour
symmetry $G_F$ (a global symmetry would give rise to a unobserved
familon Goldstone boson), and new flavon fields $\Phi$ that are charged
under $G_F$ and neutral under the Standard Model. Standard Model
matter $X_i$ is charged under both $G_F$ and $G_{SM} = SU(3) \ti
SU(2) \ti U(1)$. The superpotential is
\be
    W = \sum_{\alpha, i}
    Y_{\alpha \beta \ldots i j k} \frac{\Phi_{\alpha}}{M_X}
    \frac{\Phi_{\beta}}{M_X} \ldots X_i X_j X_k.
\ee
The flavon fields acquire vevs, spontaneously breaking the flavour
symmetry and generating the fermion mass hierarchy. The expansion
parameter for flavour symmetry breaking is ${<\Phi>}/{M_X}$.

What are the flavons in our case? $\tau_s$ and $\tau_b$ are the
only low-energy 4-dimensional fields which affect the local
geometry. As flavour symmetry breaking is controlled by the ratio
${\tau_s}/{\tau_b}$ we would like to identify the flavons with
this ratio. However this is not possible: the isometry group may
be non-Abelian - for example $SU(3)/\mbb{Z}_3$ for the $\mbb{C}^3
/ \mbb{Z}_3$ singularity - while this ratio is a real singlet and
can therefore only be in the trivial representation of the flavour
group.

The answer is somewhat subtle. The presence of the flavour
symmetry depends on an approximate isometry of the local
higher-dimensional metric. Whatever the full local metric is, we
can regard it as a perturbation on the local metric applicable to
the non-compact case.\footnote{For local models of branes at
(resolved) singularities, we again stress that by the non-compact limit
we refer to a limit which keeps the sizes of local cycles fixed
while taking the bulk volume to infinity.}
\be
    g_{MN,\textrm{local}}(y) = g_{MN,\textrm{local }, \mc{V} \to
    \infty}(y) + \delta g_{MN,\textrm{local, } \mc{V} \textrm {
    finite}}(y).
\ee
Here $g_{MN,local}$ is invariant under the flavour symmetry while
$\delta g_{MN}$ carries the form of flavour symmetry breaking. For
example, if we imagine a D7 brane wrapping the $\mbb{C}\mbb{P}^2$
of the $\mc{O}_{\mbb{P}^2}(-3)$ resolution of $\mbb{C}^3 /
\mbb{Z}_3$, then $g_{MN, local}(\mbb{C} \mbb{P}^2)$ is the
Fubini-Study metric and $\delta g_{MN}(\mbb{C}\mbb{P}^2)$ are
deformations away from the Fubini-Study metric that scale as
$\left({R_s}/{R_b} \right)^k$ for some $k$. From a four
dimensional perspective, the fields that excite $\delta g_{MN}
\neq 0$ are higher-dimensional modes of the metric, namely
Kaluza-Klein modes, and it is these modes that are charged under
the flavour symmetry.

It is not the size of the bulk \emph{per se} that breaks the local
$SU(3)/\mbb{Z}_3$ flavour symmetry, but instead the form of the
local metric. However, by Yau's theorem the full Calabi-Yau metric
is completely specified by the moduli. The complete form of the
local metric is therefore fully determined by the relative sizes
of the bulk and local cycles. The IR and UV degrees of freedom are
inter-related - although the moduli are light IR modes, the vevs
of the moduli determines the masses, vevs and mixings of the
Kaluza-Klein modes. The KK modes are the UV degrees of freedom
which act as flavons and whose vevs break the flavour symmetry. A
notable difference compared to conventional flavour models is that
while there are an infinite number of flavons in total, there are
in fact no flavons within the four dimensional effective field
theory. Once the flavons are integrated out, flavour symmetry
breaking is parametrised only by the ratio $\frac{\tau_s}{\tau_b}$
measuring the relative sizes of bulk and local cycles.

\subsection*{Isometries and Gauge Symmetries}

We would also like to relate our discussion of isometries to that
in Kaluza-Klein reduction. It is well known that in traditional
Kaluza-Klein reduction isometries of the extra dimensions give
gauge symmetries in lower dimensions, with the lower dimensional
gauge group determined by the isometry group. In our case we have
approximate local isometries, which we might expect to be
associated to approximately massless gauge bosons. We would expect
that in the limit that the isometry becomes exact, the gauge boson
should become massless. As the symmetry is geometric in nature
such fields should be geometric modes that are increasingly
massless in the large volume limit.

\vspace{2mm}
\noindent{\it Large Volume}

\vspace{2mm}\noindent Notice that the isometries are isometries
only of the local geometry and not of the full global metric.
Indeed, the isometries are not respected in any way by the bulk
geometry. The local metric has characteristic length scale $R_s$
whereas the bulk metric has characteristic scale $R_b$. Local
excitations of the geometry with length scale $R_s \lesssim R <
R_b$, corresponding to energy scales $({R_s^2 \alpha'})^{-1} < E^2
< ({R_b^2 \alpha'})^{-1}$, see the isometry as a good symmetry and
on these energy scales there is still an approximately massless
mode. However when the characteristic length scale of an
excitation reaches $R_b$, the isometry ceases to be a good
symmetry and the mode acquires a mass,
\be
    M \sim M_{KK,bulk} \sim \frac{M_{string}}{R_b} \sim \frac{M_{string}}{\tau_b^{1/4}}.
\ee
Such modes are bulk KK modes, which are
hierarchically lighter than the characteristic scale of localised
geometric modes,
\be
    \qquad M_{KK,local} = \frac{M_{string}}{R_s} \sim
    \frac{M_{string}}{\tau_s^{1/4}},
\ee
set by the curvature scale of the local geometry.

In that we wish to interpret the approximate local isometry as a
higher-dimensional gauge symmetry, this seems to single out bulk vector KK
modes as the approximately massless gauge bosons. These modes correspond to
extending the local isometry to 
transformations of the full compact space including the bulk region.
The mass scale
of these modes is hierarchically lighter than those of the local
geometry,
\be
    \frac{M_{KK,bulk}}{M_{KK,local}} \sim \frac{R_s}{R_b} \ll 1 \textrm{ for } R_b \gg 1.
\ee On the other hand the low-energy Standard Model fields have
masses bounded by the gravitino mass $m_{3/2} \sim
M_{string}/{{R_b}}^{3}$ and therefore their splittings due to the
breaking of the approximate global symmetry are much smaller than
the mass of the vector KK modes, justifying the existence of the
approximate global symmetry in the low-energy effective theory.

However there are several reasons to be cautious as to the utility of interpreting
massive bulk KK modes as gauge bosons. Most notably, these are not
symmetries for compact Calabi-Yau spaces: there is no locus in
parameter space where there actually exist massless gauge fields
with finite coupling. Regarded as a symmetry, the isometry never extends
to the bulk, which sets the compactification scale.\footnote{It is a mathematical question whether, 
irrespective of considerations of supersymmetry, there exists any metric on a space that is
topologically a compact Calabi-Yau that has the local isometry group manifest as a isometry of the entire
compact metric. These statements assume that this is not possible.}
This contrasts with for example KK reduction on $\mbb{M}_4 \ti
T^6$. The flat $T^6$ has isometries and massless gauge bosons.
However even if the isometry is not manifest and the $T^6$ metric
is highly curved, there still exists a regime of parameter space where
the metric is flat, and so every point in parameter space can be viewed as 
a (possibly large) perturbation on a limit containing
massless gauge bosons at finite coupling.

The second difficulty is that even if we do identify bulk KK modes
with approximately massless gauge bosons, there is never a gauge
symmetry in the four-dimensional effective theory. While from a
higher-dimensional perspective it may make sense to describe KK
modes as approximately massless in the limit $R_b/R_s \gg 1$,
within 4d effective field theory such a description is rather
eccentric. From the viewpoint of the 4-dimensional effective
action, the isometry is simply an approximate global symmetry of
the field theory.

We can cast this in the terminology of the list of section 2.1.
From the viewpoint of the higher-dimensional effective theory, the
vector bosons associated with approximate isometries are the bulk
KK modes. These are in the higher dimensional effective field
theory, but have couplings to the observable sector of interest
that are suppressed (as in option 3 of the list of \S2.1). By
contrast, in the low energy 4D effective theory all KK modes are
integrated out. The global symmetries can be viewed as an examples
of the second category of section 2.1. In the low energy spectrum
the approximate isometry simply acts as an approximate, possibly
nonabelian, global symmetry.

\vspace{2mm}
\noindent{\it Warped Throats}

We note that warped throats can provide similar examples of
low-energy symmetries whose origins ultimately lie in the
approximate isometries of the strongly warped regions. In this
case although the bulk breaks these isometries, the resulting
symmetry-breaking interactions only arise suppressed by powers of
the small warp factor for observers in the throat \cite{WarpSym}.
Powers of the warp factor also enter into the mass of the
corresponding KK gauge boson, but these need not be the same
powers, allowing symmetry breaking scales to be suppressed
relative to KK scales for brane modes within the throat. It should be possible to
quantify these statements using a careful an analysis of the 
low energy effective action describing
the interactions of the Standard Model fields at the tip of the
throat.

It is interesting to observe that both the above examples of approximate isometries occur
for geometries which naturally generate scales hierarchically lower than the Planck scale.
One can speculate whether this may in fact be a generic feature, and whether the systematic 
presence of approximately global geometric symmetries can be tied to the presence of new scales
hierarchically lower than the 4-dimensional Planck scale.

\section{Hyper-weakly coupled gauge groups}

We now describe another phenomenon that can arise in local brane
models, namely hyperweak interactions. While these are related to
the above discussion of approximately global flavour symmetries,
they offer novel phenomenological possibilities and are
interesting in their own right.

In string constructions it is necessary that all Standard Model
gauge factors are supported on cycles with sizes close to the
string scale. This follows from the known size and running of the
Standard Model gauge couplings,
\be
    \alpha^{-1}(m_s)  = \frac{4 \pi}{g^2(m_s)} \sim 20,
\ee
and that for a gauge group supported on (for concreteness) D7
branes wrapping a 4 cycle $\Sigma$,
\be
    \frac{4 \pi}{g^2(m_s)} = \hbox{Re}(T_s),
\ee
where $\hbox{Re}(T_s) = e^{-\phi}
\left({\hbox{Vol}(\Sigma)}/{l_s^4} \right)$.

However, generally we expect additional branes in the
compactification beyond simply those of the Standard Model,
leading to additional hidden sectors. One particularly interesting
possibility is the existence of bulk branes that wrap cycles
supported on the entire Calabi-Yau. On dimensional grounds, the
size of such a 4-cycle is $\tau_b \sim \mc{V}^{2/3}$ which for
LARGE volume leads to extremely weak gauge couplings. Numerically,
for the case of $\mbb{P}^4_{[1,1,1,6,9]}$ (this Calabi-Yau has
featured prominently in studies of the LARGE volume scenario)
$\tau_b = 162^{-1/3} \mc{V}^{2/3} \sim 0.18 \mc{V}^{2/3}$.
Requiring a volume of $\mc{V} \sim 10^{14}$ to generate TeV soft
terms, a bulk brane will generate a gauge coupling
\be
    \frac{4 \pi}{g^2} = \tau_b \sim 0.18 \ti \mc{V}^{2/3}
    \sim 4 \ti 10^{8},
\ee
corresponding to $g \sim 2 \ti 10^{-4}$. This produces a force
characterised by exceptionally weak coupling. This geometric
picture is illustrated in figure \ref{weakgroup}.
\FIGURE{\epsfig{file=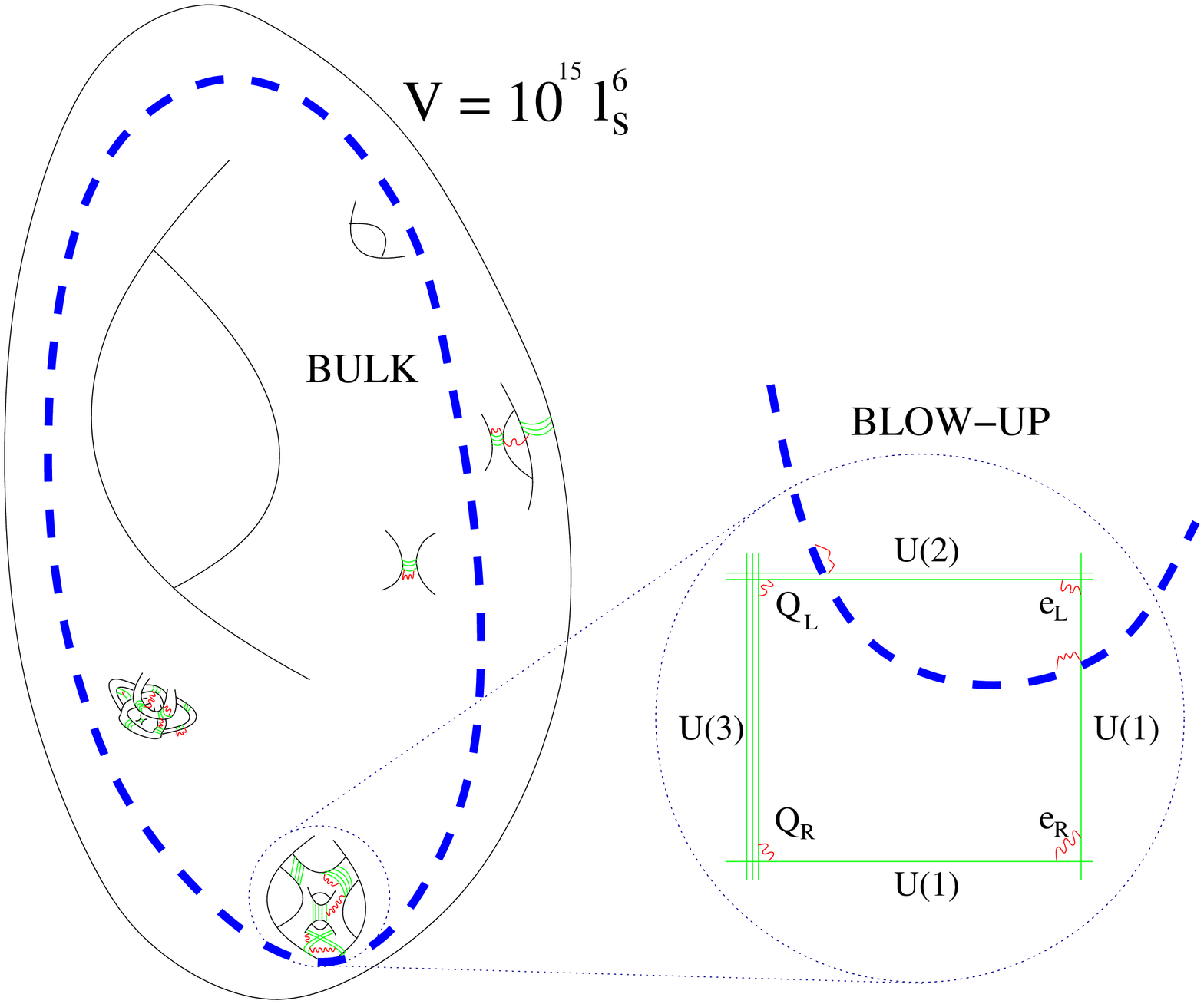,width=\textwidth}
        \caption[p]{\footnotesize{An illustration of the presence
        of a brane, that wraps a bulk cycle of the Calabi-Yau and
        carries a very weakly coupled gauge group, intersecting with the
        Standard Model branes.}}
    \label{weakgroup}}

We therefore see that the existence of a hyper-weak gauge boson is
a general feature of LARGE volume models. The phenomenology will
depend on the mass scale of such a hyper-weak boson and how it
mixes with the Standard Model photon. If the hyper-weak boson is
entirely decoupled from the Standard Model then it will play
little role. However we note that in phenomenological models of
branes at singularities, such as those developed in
\cite{hepth0005067}, there often exist flavour D7s, which wrap
bulk cycles but directly intersect the local branes carrying the
Standard Model gauge group.

Let us enumerate possible mass scales for the hyper-weak gauge boson. If the
hyperweak U(1) is coupled to the quark sector, then the condensate
$ \langle \bar{q} q \rangle \sim \Lambda_{QCD}^3$ will spontaneously break the
U(1), leading to
\be \label{eq:mZprime_keV}
    m_{Z'}\sim g\Lambda_{QCD} \sim 10^{-4} \times 100 \;\textrm{MeV}
    = \mc{O}(10) \; \textrm{keV}.
\ee
If the breaking arises from the MSSM down-type Higgs, then
assuming $\tan \beta \sim 10$ the expected mass scale is
\be \label{eq:mZprime_mass2}
    m_{Z'} \sim 10^{-4} \ti 20 \, \hbox{GeV} \sim 1 \, \hbox{MeV}.
\ee
Finally, if $U(1)_{HW}$ is broken by the Standard Model Higgs, or
the MSSM up-type Higgs, the expected mass scale is
\be \label{eq:mZprime_mass1}
    m_{Z'} \sim 10^{-4} \ti 246 \, \hbox{GeV} \sim
    \mc{O}(10) \, \hbox{MeV}.
\ee
The hyper-weak boson may also be coupled to hidden Standard
Model-like sectors from different regions of the bulk, generating
$m_{Z'} \sim (10^{-4} \to 10^{-5}) \ti v_{hidden}$. If
$v_{hidden}$ is the weak scale - which is the scale of
supersymmetry breakdown and the scale at which soft terms are
generated - this again gives a mass for the hyper-weak boson of
$\sim 10 \hbox{MeV}$.

We now discuss the phenomenology and current experimental bounds
on hyper-weak interactions with different boson masses.  In
general, massless hyper-weak bosons are tightly constrained by fifth
force experiments. For MeV scale vector boson, our discussion
mainly follow a series of papers by Fayet
\cite{Fayet1980:PLB95, Fayet:2006xd, hepph0607318,  Fayet:2007ua}.  We will see that
low energy precision observables, such as $g_e -2 $ and $g_{\mu} - 2$
 and $e-\nu$ scattering, constrains
the coupling of the lepton sector to the hyperweak U(1).
Axion-like decay of the $\Upsilon(b\bar{b})$ and $\psi(c\bar{c})$
mesons provide tight constraints on the coupling of the quarks,
whereas atomic parity violation constrain coupling product of the
quark and electron. We shall see that many of these bounds are at
comparable orders of magnitude to the generic expectation of
the size of the hyper-weak coupling in the LARGE volume scenario. We also
discuss the possibility of having light dark matter residing
on the bulk brane.  Finally, the experimental constraints on
hyperweak gauge bosons
with electroweak or higher scale masses are briefly
reviewed.  These constrains mainly come from electroweak precision
observables and collider searches and as expected
are rather mild.

\subsection{Mixing between the visible sector and the hyper-weak U(1)}

\FIGURE{\epsfig{file=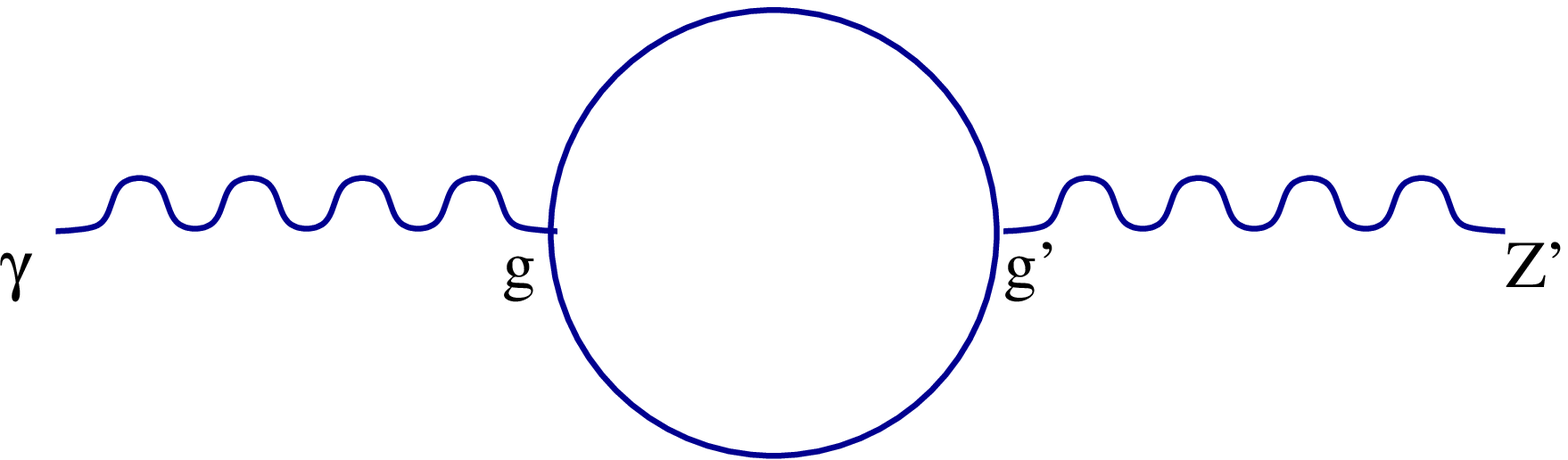,width=.5\textwidth}
        \caption[p]{\footnotesize{The mixing between two
 $U(1)$'s induced by loops of particles charged
        under both.}}
    \label{mixing}}

In explicit string cosntructions of quasi-realistic models it is
very common to have matter sectors living on different branes.
There are usually several $U(1)$ gauge fields that have to be
properly normalised. Even if different $U(1)$'s can be
normalised to be orthogonal, loop corrections will induce mixing
between different $U(1)$'s.   These will then induce an effective
charge to the Standard Model fields with respect to an otherwise
hidden $U(1)$. The phenomenology of light weakly coupled gauge
bosons has been studied in some detail in \cite{hepph0607318,
Fayet:2007ua, steve, Abel:2008ai, ringwald, millicharge}, and depends in an
important way on the relative size of the mixings and mass terms.
Consider, for instance, a system of two 4D gauge bosons,
$A^a_\mu$, subject to the following Lagrangian
\be
    {\cal L} = - \frac14 \, Z_{ab} A^a_{\mu\nu} A^{b\,\mu\nu}
    - \frac12 M^2_{ab} A^a_{\mu} A^{b\,\mu} - j^\mu_a A^a_\mu \,,
\ee
where $A^a_{\mu\nu} = \partial_\mu A^a_\nu - \partial_\nu A^a_\mu$
and $j^\mu_a$ denote the currents to which each gauge boson
couples. Positive kinetic energy requires the matrix $Z_{ab}$ to
be positive definite.

The kinetic and mass terms may be diagonalised in the usual way,
through the transformation
\be
    A^a_\mu = {N^a}_b \, {\cal A}^b_\mu
    = {\left(Z^{-1/2}\right)^a}_c \, {R^c}_b \,
    {\cal A}^b_\mu\,,
\ee
where
\be
    R = \left(%
    \begin{array}{rrr}
    \cos\theta && -\sin\theta \\
    \sin\theta && \cos\theta \\
    \end{array}%
    \right) \,,
\ee
denotes the two-by-two rotation matrix that diagonalises the
symmetric matrix defined by $\hat M^2 = Z^{-1/2} M^2 Z^{-1/2}$.

Explicitly, taking without loss of generality
\be
    Z = \left(%
    \begin{array}{ccc}
    1 && \lambda \\
    \lambda && 1 \\
    \end{array}%
    \right) \qquad \hbox{and}\qquad
    M^2 = \left(%
    \begin{array}{cc}
    \ma^2 & \mu^2 \\
    \mu^2 & \mb^2 \\
    \end{array}%
    \right)
\ee
we have
\be
    Z^{-1/2} = \sec 2\alpha \left(%
    \begin{array}{rrr}
    \cos\alpha && -\sin\alpha \\
    -\sin\alpha && \cos\alpha \\
    \end{array}%
    \right) \,,
\ee
where the angles $\theta$ and $\alpha$ are given by
\be
    \sin 2\alpha = \lambda \qquad \hbox{and} \qquad
    \tan 2\theta = \frac{2\mu^2 - (\ma^2 + \mb^2)
    \lambda}{(\ma^2 - \mb^2) \sqrt{1-\lambda^2}} \,.
\ee
The corresponding mass eigenvalues then are
\be
    M^2_{\pm} = \frac{\ma^2 + \mb^2 - 2 \lambda \mu^2
    \pm \Delta}{2(1 - \lambda^2)} \,,
\ee
where
\be
    \Delta^2 = (\ma^2 - \mb^2)^2 + 4\mu^4 - 4\lambda \mu^2 (\ma^2
    + \mb^2) + 4\lambda^2 \ma^2 \mb^2 \,.
\ee

The final Lagrangian is
\be
    {\cal L} = -\frac14 \, {\cal A}^a_{\mu\nu} {\cal A}^{\mu\nu}_a
    - \frac12 \, \left( M_+^2 {\cal A}^+_\mu {\cal A}^\mu_+
    + M_-^2 {\cal A}^-_\mu {\cal A}_-^\mu \right)
    + {N^a}_b j^\mu_a {\cal A}^b_\mu \,,
\ee
where the combined redefinition is
\be
    N = \frac{1}{\sqrt{1-\lambda^2}} \left(%
    \begin{array}{ccc}
    \cos(\theta+\alpha) && -\sin(\theta + \alpha) \\
    \sin(\theta-\alpha) && \cos(\theta - \alpha) \\
    \end{array}%
    \right) \,.
\ee
Several limiting cases are of particular interest.
\begin{itemize}
\item In the absence of kinetic mixing, $\lambda \to 0$, we have
$\alpha \to 0$ and so $N \to R$ reduces to the usual rotation,
with mixing angle $\tan 2\theta = 2\mu^2/(\ma^2 - \mb^2)$.
\item If $\ma^2 = \mu^2 = 0$, then one of the gauge fields is
massless, $M_-^2 = 0$, while the other has mass $M_+^2 =
\mb^2/(1-\lambda^2)$ and $\sin 2\theta = \sin 2\alpha = \lambda$.
In this case only the massive vector acquires a coupling to both
currents
\be \label{eq:masslessmassivemixing}
    {\cal L}_{\rm int} = j_1^\mu {\cal
    A}^-_\mu + \frac{1}{\sqrt{1-\lambda^2}}
    \left( j_2^\mu - \lambda
    j_1^\mu \right) {\cal A}^+_\mu \,.
\ee
\item If the entire mass matrix vanishes, $M^2_{ab} = 0$, then we
may take $R = 1$ and so $\theta = 0$. In this case $N = Z^{-1/2}$
and so both gauge bosons in general couple to both currents
\be\label{eq:masslessBoson}
    {\cal L}_{\rm int} = \frac{\cos\alpha}{\sqrt{1-\lambda^2}}
    \left( j_1^\mu {\cal A}^-_\mu
    + j_2^\mu {\cal A}^+_\mu \right)
    - \frac{\sin\alpha}{\sqrt{1-\lambda^2}}
    \left( j_1^\mu {\cal A}^+_\mu
    + j_2^\mu {\cal A}^-_\mu \right) \,,
\ee
with $\sin^2\alpha = \frac12 \left[ 1 - \sqrt{1-\lambda^2}
\right]$ and $\cos^2\alpha = \frac12 \left[ 1 + \sqrt{1 -
\lambda^2} \right]$ (and so $\sin 2\alpha = \lambda$, as above).
\end{itemize}

These limits show that the nature of the bound on new gauge bosons
depends in an important way on whether or not the boson masses
play a significant role. Since the photon mass is known not to be
larger than $3 \times 10^{-36}$ GeV \cite{PDG} (see also, however,
\cite{ADG}), only the latter two categories are relevant for
photon energies greater than this (or for wavelengths smaller than
a few kpc). Within this range, it is eq.~\pref{eq:masslessBoson}
which applies for photon-mixing applications when the wavelengths
involved are much longer than the hyper-weak gauge boson's Compton
wavelength. Otherwise it is eq.~\pref{eq:masslessmassivemixing}
that applies. The second regime of application of the above
formulae is to mixings between the hyper-weak boson and the Standard
Model $Z$ boson (or the hypercharge gauge boson, $B_\mu$), as is
appropriate for applications to bounds coming from particle
accelerators.

\subsection{Bounds on light vector bosons}

If the hyper-weak bosons are sufficiently light, then their
mixings with photons are described at low energies either by
eq.~\pref{eq:masslessmassivemixing} or
eq.~(\ref{eq:masslessBoson}), depending on the size of the
hyperweak boson mass compared with the wavelengths present in the
experiments. In this case there are two types of bounds: those
constraining the existence of a new light boson coupled to
ordinary electrically charged particles; and those constraining
the couplings of the photon to exotic millicharged particles
\cite{holdom}, that start life as particles coupled only to the
hyperweak boson. Of course, the millicharge constraints are only
relevant if such new particles exist in addition to the low-energy
hyperweak gauge boson.

\subsubsection*{Direct Bounds on Vector Bosons}

By far the strongest constraints arise if the couplings of the
light gauge boson can violate the principle of equivalence, and if
the mass is small enough that the range of the new hyperweak force is
macroscopically large. In this case very strong bounds arise,
coming from fifth force experiments on Earth
\cite{Adelberger:2003zx}, and from other tests of General
Relativity \cite{PDG,GRTests}. For example, lunar laser-ranging
limits the difference between the acceleration of Earth and Moon
towards the Sun to be \cite{PDG}
\be
    \frac{\Delta a}{a} = (-1.0 \pm 1.4) \times 10^{-13} \,,
\ee
and so if the mass of the hyperweak boson, $U$, satisfies $m_U <
10^{-18}$ eV, then
\be
    \frac{\left|Q_S ( Q_E/M_E-Q_M/M_M)\right|}{4\pi GM_S}
    \lsim 10^{-13}
    \,,
\ee
where $M_I$ and $Q_I$ for $I=E,M,S$, respectively denote the mass
and hyperweak charge for the Earth, Moon and Sun. If the hyperweak
boson were to couple to lepton number with strength $g$, for
instance, then to good approximation we have $Q_I/M_I = g
N_{eI}/(m_n N_{nI}) = g X_I/m_n$, where $N_{eI}$ and $N_{nI}$
count the total number of electrons and nucleons in object `$I$'
and $m_n$ is the nucleon mass. $X_I$ here denotes the proton
fraction, $X_I = N_{pI}/N_{nI} = N_{eI}/N_{nI}$, and so, being
made mostly of hydrogen, for the Sun $X_S \approx 1$. In this case
the bound on $g$ is very strong:
\be
    g \lsim 10^{-6} \left(
    \frac{Gm_n^2}{|X_E-X_M|}\right)^{1/2} \sim
    \frac{10^{-25}}{\sqrt{|X_E - X_M|}} \,.
\ee

If, however, the hyperweak boson couples only to ordinary matter
through its mixing with the photon, then its coupling is strictly
proportional to the electric charge, $Q_{\rm em}$, of the source.
In this case the direct bounds on the existence of new light
bosons are weaker than above because of the difficulty of directly
detecting the presence of the hyperweak-photon in the presence of
the huge background presented by ordinary electromagnetic
interactions. In the absence of millicharged particles the best
direct bounds then come from the failure to observe
photon/hyperweak-photon oscillations
\cite{Abel:2008ai,photonoscillations} and from tests of the
inverse square form of the Coulomb interaction
\cite{photonyukawa}, if the mass of the hyperweak boson, $U$, lies
in the range $10^{-9}$ eV $< m_U < 10^5$ eV relevant for
experiments on Earth or oscillations between the Earth and the
Sun.

The resulting bounds constrain the kinetic mixing $|\lambda| \lsim 10^{-8}$ if $m_U
\sim 10^{-6}$ eV, but extend down to $|\lambda| \lsim 10^{-13}$
for hyperweak masses of order $m_U \sim 10^2$ eV. The bounds
deteriorate to $|\lambda | \lsim 10^{-5}$ once $m_U \lsim 10^{-9}$
eV or $m_U \gsim 10^5$ eV. If $\lambda$ is generated by loops of
4D fermions of mass $m$ carrying both a unit of ordinary and
hyperweak charges, then $\lambda \sim [eg^V/(4\pi)^2] \log
(m/\mu)$, where $e$ and $g^V$ are the strengths of the
electromagnetic and vector hyperweak couplings and $\mu$ is an
appropriate renormalization point.\footnote{See ref.~\cite{steve}
for calculations of $\lambda$ in more complicated stringy and
higher-dimensional frameworks.} In this case a bound like
$|\lambda| \lsim 10^{-8}$ would correspond to $g^V \lsim 10^{-6}$.

\subsubsection*{Bounds on Millicharged Particles}

Another strong class of bounds exists if the hyperweak gauge boson
is effectively massless on the scale of the experiments of
interest, and there also are other exotic light particles that
initially couple only to the hyper-weak boson. In this case
eq.~\pref{eq:masslessBoson} implies that these new exotic
particles acquire miniscule couplings to the ordinary photon
\cite{holdom,millicharge,Abel:2008ai}. The absence of evidence for
such particles in astrophysics and cosmology excludes mixings of
order $|\lambda| \lsim 10^{-14}$ for $m_U \lsim 10^4$ eV, while
Earth-based experiments require $|\lambda| \lsim 10^{-4}$ for $m_U
\lsim 10^6$ eV. Limits quickly deteriorate for masses much larger
than this.

\subsection{Massive vector bosons}

Since the mixings of massive vector bosons with photons tends not
to generate mini-charged particles, the nature of the constraints
obtained in this case can be weaker.  Since precision electroweak
measurements test the form of fermion couplings to the photon and
Z boson, they provide limits on the existence of hidden sector
vector bosons that mix with these particles.

Depending on the brane model constructed, the hyperweak U(1) can
either couple to the lepton or quark sectors only, or to both
sectors simultaneously.  If the bulk and the SM branes do not
intersect at all, the hyperweak U(1) could still couple to the
visible sector via the NS-NS $B_2$ and R-R forms. The latter leads
to kinetic and mass mixings, and is discussed in detail in
\cite{Abel:2008ai}.  Here we focus on the case when the bulk and
the SM branes intersect.

\begin{table}[ht!]
\centering
\begin{tabular}{|c|c|cr|}
\hline
Coupling & Bound & Experimental measurement & \\
\hline
$g_e^V$ & $10^{-4} m_{U}$ & $g_e-2$ &\cite{Fayet1980:PLB95}\\
$g_e^A$ & $5 \, 10^{-5}m_{U}$ & $g_e-2$& \cite{Fayet1980:PLB95}\\
$g_{\mu}^V$ & $10^{-3}$ &  $g_{\mu}-2$&\cite{Fayet:2007ua}\\
$g_{\mu}^A$ & $5 \, 10^{-6} m_{U}$ &  $g_{\mu}-2$&\cite{Fayet:2007ua}\\
$|g_e g_{\nu}|$ & $10^{-11}m^2_{U}$ & $\nu-e$ scattering &\cite{Boehm:2003hm} \\
\hline
$g_{c(b)}^A$ & $10^{-6}m_{U}$ & $B(\psi(\Upsilon)
\to \gamma + \textrm{invisible})$ &\cite{hepph0607318}\\
\hline
$|g_e^A g_q^V|$ & $10^{-14}m^2_{U}$ & atomic parity violation
&\cite{Bouchiat:2004sp} \\
\hline
\end{tabular}
\caption{Approximate bounds on coupling of standard model fermions
to a MeV scale hyperweak vector boson $U$.  The superscript V(A)
denotes vector(axial) couplings.  If no superscript is present,
the coupling correspond to $\sqrt{(g^V)^2+(g^A)^2}$.  The
subscript labels the particle to which the bounds should be
applied.  The mass $m_{U}$ is in units of MeV.}
\label{tab:MeVbound}
\end{table}

For a vector boson U of $\mc{O}(10)\textrm{MeV}$, a number of
experimental bounds exist on various couplings of the SM fermions
to this `hidden' sector boson.  A collection of such bounds
adapted from a series of papers by Fayet is displayed in
table~\ref{tab:MeVbound}.\footnote{For our present purpose, it is
not necessary to detail the $\mc{O}(1)$ coefficients, as we are
more interested in the order of magnitude estimate.} In general
the vector and axial coupling bounds are different, and their
scaling with the U boson mass $m_{U}$ depends on the scale of the
process involved.  For instance, for the calculation of the
anomalous magnetic moment of a fermion $f$, both the vector and
axial contributions can be cast into the form \be \delta
a^{V(A)}_f \sim \Big(g^{V(A)}\frac{m_f}{m_U}\Big)^2
L^{V(A)}\Big(\frac{m^2_f}{m^2_U}\Big), \ee where $L^{V(A)}$ are
loop functions.  The axial coupling $g^A$ scales as $m_{U}$ due to
the enhancement of order $m^2_{f}/m^2_{U}$ from the longitudinal
part of the U propagator, and is reflected by the fact that
$L^{A}$ can be approximated by a constant.  For the vector part,
when $m_{U} \gg m_f$, $L^{V}\sim \textrm{const.}$, hence again we
have $g^V\sim m_{U}$. On the other hand, for $m_{U} \ll m_f$, the
situation is analogues to QED contributions to $\delta a$, and we
have instead $\Big(\frac{m_f}{m_U}\Big)^2 L^{V}\sim
\textrm{const.}$

It should be noted that this set of bounds may be considered as
applying separately to leptons or quarks, or to both.  Which of these
couplings are relevant is a model dependent question.
However it is interesting to observe that the bounds are
comparable to the estimated hyperweak coupling in the LARGE volume
scenario with $m_{U}\sim \mc{O}(10)\textrm{MeV}$.\footnote{An
exception is the bound from atomic parity violation which
constrains the quark and electron coupling product.  This could
have implications on model building.}

For a vector boson of $\mc{O}(10)$ keV discussed before
(\ref{eq:mZprime_keV}), the axial coupling bounds are still valid,
implying very stringent bounds on the axial couplings of
$\mc{O}(10^{-7})$ or below.  The vector coupling bound for the
muon remains the same, whereas the one for the electron is reduced
to $\mc{O}(10^{-5})$.  It thus appear unlikely that a keV scale
vector boson would be compatible with the LARGE volume models
under consideration.

The existence of a hidden brane with a weakly coupled U(1) also
leads to the natural possibility of having dark matter ($\chi$)
residing in the hidden sector.  An interesting scenario of having
light dark matter (LDM) of $\mc{O}(10)$ MeV was discussed in
\cite{Fayet:2006xd, Boehm:2003hm, Borodatchenkova:2005ct}.  In these works, the
couplings of the LDM and those of the SM particles to the extra
U(1) are un-related.  However in the LARGE volume scenario, both
type of couplings are expected to be of $\mc{O}(10^{-4})$.  Using
the result that the total LDM annihilation cross section at freeze
out should be roughly equal 4 to 5 pb, a relation obtained in
\cite{Fayet:2006xd} implies in general
\be
    |f_{\chi}||f_{SM}| \simeq 10^{-6}
    \frac{|m^2_{U}-4m^2_{\chi}|}{m_{\chi}(1.8\textrm{MeV})},
\ee
where $f_{\chi}$ and $f_{SM}$ are generic couplings of the LDM
and SM particles to the hyperweak U(1).  We see that the couplings
involved in the hyperweak sector is close to the ball park leading
to efficient LDM annihilation into SM particles.

For a heavy vector boson with mass $m_{V}$ of the electroweak
scale or above, limits can be obtained from electroweak precision
observables and collider $Z'$ searches.  Constraints from
electroweak precision observables generally limit $m_{V}$ to be
at least a few hundred GeV, and upper bound on kinetic mixing of
$\mc{O}(10^{-3})$ \cite{Erler:1999ub}.  There is essentially no
constraints on a massive vector boson with mass above a TeV
\cite{Wells:2008xg}.  The agreement between SM predictions and
LEP-II measurements on $e^+e^- \to f\bar{f}$ implies either $m_{V}
> 209\textrm{GeV}$, and/or that its coupling to $e$ and $f$ be smaller
than $\mc{O}(10^{-2})$.  Tevatron searches have also set limits on
$Z'$ couplings to u- and d- quarks \cite{Abulencia:2005nf}, with
coupling limits again of $\mc{O}(10^{-2})$ which are relaxed
approximately logarithmically with $Z'$ mass.

\section{Conclusions and Outlook}

In this note we have analysed general aspects of continuous global
symmetries (abelian and non-abelian) in string theory and their
implementation in local models within the context of LARGE volume
string compactifications. Assuming that a Standard Model-like
theory is engineered through an appropriate local combination of
branes, we have identified two interesting phenomenological
features. The first is that the field theory couplings and
interactions are expected to be determined by approximate
continuous global flavour symmetries, which are inherited from the
approximate isometries of the local metric. While consistent with
precise statements on global symmetries in string theory
\cite{BanksDixon}, this is a counterexample to the folk theorem
that continuous global symmetries are not relevant to effective
field theories derived from string compactifications. A particularly attractive feature
is that it may allow both the fermion mass and
weak hierarchies to be simultaneously generated by the LARGE
volume. The natural expansion parameter for the breaking of the
flavour symmetry is the inverse radius of the compact space,
$l_s/R_b \sim 0.01$. While on the surface such a parameter is very
attractive for explaining the fermion mass hierarchy, more work is
necessary to carefully determine the correct powers of this
parameter that will enter the expansion in a realistic model.

The second interesting feature of such models is the possible
existence of new hyper-weakly coupled gauge groups under which
Standard Model matter is charged. Such groups would have
$\alpha_{bulk} \sim 10^{-9}$ and are associated with branes that
wrap bulk cycles but have local intersections with the Standard
Model branes. If broken by bifundamental matter acquiring a
weak-scale vev - for example the Higgs fields - the gauge bosons
of such groups would acquire masses of $m_{Z'} \sim 10 \hbox{MeV}$
through the Higgs mechanism. While difficult to discover at
high-energy accelerators, the existence of such light,
weakly-coupled gauge bosons may be accessible to low-energy
precision experiments.

Let us finish with some general remarks regarding cosmological
implications. The nonexistence of exact global symmetries has
important implications for several cosmological scenarios. In
particular there are (non) topological defects that require a
global symmetry to be present - for example textures, global monopoles, global
strings or semilocal strings \cite{paul}. If all symmetries are gauged, we may be
inclined to conclude that these defects should not be present as a
a general {\it prediction} of string theory. However, based on the
logic of this paper we may wonder whether approximate low-energy global
symmetries may be sufficient
for these defects to play a cosmological role.

\begin{enumerate}
\item{\it Global strings and monopoles.} Cosmic strings and
monopoles can arise from the breaking of global or local
symmetries and the topological conditions for their existence are
the same for both global and local symmetries. However the
physics is very different. For instance the force among global
monopoles/strings is independent of the distance and they
annihilate more efficiently than the local ones, and the presence of a
Goldstone mode provides a channel for radiation that is not
available in the local case.
Even though exact global monopoles and strings may not be allowed in string
theory,
we can still think about approximate global defects and estimate the
effect these would have in cosmology. As described in the paper we
expect the would be Goldstone modes to be bulk modes with mass
at the KK scale.
While detailed
physical implications of this scenario are beyond the scope of
this article, it is clear that the quality of the approximate symmetry
will play an important role.

\item{\it Semi-Local Strings.} Semi local strings are formed when
a product $G_{global}\times G'_{local}$ is broken to $H_{global}$
and the vacuum manifold is simply connected. In the simplest case
$G_{global}=SU(2)$ and $G'_{local}= U(1)_{local}$ and
$H_{global}=U(1)_{global}$. As the full local $U(1)$ is broken
cosmic strings are expected \cite{semilocal}. Again,
semi-local strings may exist due to the approximate global
symetries, with stability properties determined by the
amount of breaking. For a recent discussion on semi local strings
and their potential role in string theoretical models see
\cite{linde}.

\item{\it Textures.} In the standard classification of topological
defects, if the vacuum manifold ${\cal{M}}$ is such that the third
homotopy group is non-trivial $\pi_3({\cal{M}}) \neq I$
textures are induced, corresponding to a non trivial winding of
the field once the points at infinity in our 3D space are
identified (turning it into an $S^3$). In contrast to other
defects the scalar field is always in the vacuum manifold and
therefore only gradients of the scalar field conribute to the
energy. These would be naturally cancelled by a non-trivial
configuration for the gauge field in the covariant derivative and
therefore mostly global textures have been considered for
cosmological implications. By Derrick's theorem it is clear that a
texture is unstable to collapse, but in an expanding universe it
stretches until it comes inside the horizon when it starts to
shrink. As for other defects, the gravitational field of textures
can distort the isotropy of the CMB. Their effects have been
recently studied in \cite{neil} in order to explain the cold spot
observed in the CMB. Again, the absence of global symmetries in
string theory would indicate that global textures are not
possible. But an approximate global symmetry originating from a
gauge symmetry may still allow the formation of early universe
textures. For instance for a light gauge field $A_\mu$ with mass
$m_A$, a local texture of size $L\ll 1/m_A$ will behave as a
global texture.

\end{enumerate}

Overall, it is interesting that both Abelian and non-Abelian continuous
global symmetries can be present in string theory, albeit in an
approximate way. In the examples we have studied the presence of the approximate global symmetry
is tied to extra-dimensional geometries that naturally give rise to hierarchies.
Such approximate global symmetries can have important phenomenological and
cosmological implications. It will be worthwhile both to further investigate their implications
and to study explicit realisations in semi-realistic local D-brane models, and we hope to return to this topic in future.

\acknowledgments{} We are grateful for discussions with B. Allanach, R.
Brustein, B. Dolan, N. Dorey, J. Garriga, J. Gauntlett, J.
Polchinski, G. Ross, E. Sezgin, P. Shellard, N. Turok and M.
Williams. CB is supported in part by research funds from
the Natural Sciences and Engineering Research Council (NSERC) of
Canada, the Killam Foundation and McMaster University. Research at
the Perimeter Institute is supported in part by the Government of
Canada through NSERC and by the Province of Ontario through MRI.
He thanks the Centre for Theoretical Cosmology (CTC) at Cambridge University for
hospitality while part of this work was done.  JC is funded by
Trinity College, Cambridge. He also thanks the University of Texas
at Austin for hospitality while part of this work was carried out
and was supported in part by NSF Grant PHY-0455649.
CHK is supported by a Hutchison Whampoa Dorothy Hodgkin
Postgraduate Award. LYH
thanks the Gates Cambridge Trust and ORSAS UK for financial support.
AM is supported by STFC. FQ is suppoerted by STFC and a Royal Society
Wolfson Award. He acknowledges the Mitchell family and the
organizers of the Cook's Branch meeting 2008 at Texas A\&M for hospitality.

\end{document}